# Dynamics of surface solitons at the edge of chirped optical lattices

Yaroslav V. Kartashov,[1] Victor A. Vysloukh,[2] and Lluis Torner[1]

[1]ICFO-Institut de Ciencies Fotoniques, and Universitat Politecnica de Catalunya, Mediterranean Technology Park, 08860 Castelldefels (Barcelona), Spain

[2]Departamento de Fisica y Matematicas, Universidad de las Americas – Puebla, Santa Catarina Martir, 72820, Puebla, Mexico

We address soliton formation at the edge of chirped optical lattices imprinted in Kerr-type nonlinear media. We find families of power thresholdless surface waves that do not exist at other types of lattice interfaces. Such solitons form due to combined action of internal reflection at the interface, distributed Bragg-type reflection, and focusing nonlinearity. Remarkably, we discover that surfaces of chirped lattices are soliton attractors: Below an energy threshold, solitons launched well within the lattice self-bend toward the interface, and then stick to it.

Nonlinear surface waves localized at the very interface between two media exhibit unique properties and might be potentially useful for practical applications such as surface characterization, optical sensing, and switching [1,2]. Many nonlinear surface waves exist only above a power threshold. Recently it was predicted and experimentally observed that nonlinear self-confinement of light beams near the edge of waveguide array with focusing nonlinearity leads to the formation of discrete surface solitons for high enough input powers [3,4]. Interfaces between lattices with defocusing nonlinearity and uniform media support surface gap solitons [5,6] that have been experimentally observed [7,8] and surface kink solitons [9]. The studies of surface waves at lattice interfaces were extended to quadratic [10], saturable [11], nonlocal nonlinear materials [12] and to interfaces between complex periodic structures [13,14]. Polychromatic surface solitons, and spectrally-selective attraction or repulsion at the surface was studied in Refs. [15,16]. In all these settings surface solitons are bound to the interface because their propagation constants belong to forbidden gaps of the periodic lattice spectrum; thus, nonlinearity is required to produce the necessary shift of the propagation constant. A spatial modulation of lattice parameters (i.e., its period or strength) can change the conditions of surface soliton existence dramatically since modulation modifies the band-gap lattice structure (if modulation is weak) or completely destroys it (if it is strong). The possibilities thus opened for soliton control are numerous [17-19].

In this paper we study, for the first time to our knowledge, surface solitons that bifurcate from the fundamental linear modes supported by the interface of a uniform medium and a lattice with linearly chirped frequency. Such linear modes feature unstaggered phase structure in contrast to fundamental linear modes encountered at the interface of periodic layered media [20,21]. In contrast to solitons supported by the interfaces of periodic focusing waveguide arrays, we find that the formation of solitons at chirped lattice interfaces is power thresholdless. We also reveal that the interface of the chirped lattice acts as soliton attractor.

We consider the propagation of a laser beam along the $\xi$-axis near the interface of a semi-infinite period-chirped lattice imprinted in a focusing nonlinear medium. Light propagation is described by the nonlinear Schrödinger equation for the dimensionless field amplitude $q$:

$$i\frac{\partial q}{\partial \xi} = -\frac{1}{2}\frac{\partial^2 q}{\partial \eta^2} - |q|^2 q - pR(\eta)q, \qquad (1)$$

where $\eta, \xi$ stand for the normalized transverse and longitudinal coordinates, respectively; $p$ is the lattice modulation depth; the function $R(\eta) = 0$ at $\eta < 0$ and $R(\eta) = 1 - \cos[\Omega(\eta)\eta]$ at $\eta \geq 0$ describes the profile of the semi-infinite lattice with the linearly increasing spatial frequency $\Omega(\eta) = \Omega_0(1 + \alpha\eta)$; the parameter $\alpha > 0$ stands for the rate of linear frequency chirp. Such lattice interfaces can be fabricated by different methods, e.g. by etching ridge waveguides with properly adjusted widths onto a semiconductor substrates [4] or by titanium in-diffusion on the surface of the photorefractive material [7,8]. A potential alternative to produce modulated lattice interfaces in suitable photorefractive crystals is optical lattice induction combined with erasure of the part of complex interference pattern of several red-light beams having different propagation angles by green-light background illumination. We search for stationary solutions of Eq. (1) in the form $q(\eta,\xi) = w(\eta)\exp(ib\xi)$, where $w$ is the real function and $b$ is the propagation constant.

A unique feature of chirped lattice interfaces is that they can support linear guided modes if $\alpha > 0$. In the linear limit the energy exchange rate between the incident plane wave with transverse wavenumber $k_0$ and Bragg-scattered wave with wavenumber $-k_0$ is determined by the Fourier lattice spectrum $R_k(k) = (4/\pi\alpha\Omega_0)^{1/2} \cos[(\Omega_0^2 + k^2)/4\alpha\Omega_0 - \pi/4]$ taken in the point $k = 2k_0$ [22]. Note, that the function $|R_k(k)|$ has a flat plateau in the vicinity of $k = 0$, provided that the chirp rate is $\alpha = \Omega_0[\pi(1+4m)]^{-1}$ for $m = 0,1,2...$ The width of this flat reflection band is given by $\Delta k = 2(2\pi\alpha\Omega_0)^{1/2}$ and thus this band is suitable for the formation of linear surface waves propagating along the interface. The distance of energy exchange between the Fourier component of the surface wave with normalized transverse wavenumber $k_0$ (propagating into the lattice depth) and the Bragg-reflected wave $-k_0$ is given by $L_e = \pi/[2p|R_k(2k_0)|]$. If the surface wave spectrum belongs to the band $(-k_0, k_0)$, then the maximal length of transverse localization $L_t$ can be estimated as $L_t = k_0 L_e$ taking into account that in normalized variables $k_0$ is the tangent of the propagation angle. The transverse localization length diminishes with increase of $p$ (stronger guiding and consequently more pronounced near-the-surface localization) and grows with the lattice carrying frequency $\Omega_0$.

Figure 1 illustrates the basic properties of the linear modes supported by the chirped lattice interfaces. We found such modes numerically by solving the eigenvalue problem that is obtained from the linear version of Eq. (1). The panel (a) shows the normalized propagation constant $\delta = (b-p)/p$ of the lowest-order nodeless surface mode as a function of lattice modulation depth $p$ for different values of the lattice frequency $\Omega_0$. With increase of $\Omega_0$ the inflection point shifts toward higher values of $p$. The profiles of the corresponding linear surface modes are depicted in Fig. 1(b). The key feature of the distributions is that increasing the lattice modulation depth $p$ enhances the near-surface mode localization, with the absolute mode maximum gradually shifting from the lattice interior to the nearest to interface lattice channel. Thus, with increase of $p$ the collective excitation of strongly coupled lattice channels is gradually replaced by self-trapping in a single surface channel. Figure 1(c) shows the monotonic growth of the critical value of lattice modulation depth $p_{cr}$ (for $p \geq p_{cr}$ the mode maximum is located in the first guiding channel, while for $p < p_{cr}$ it shifts into the lattice interior) with increase of the carrying lattice frequency $\Omega_0$. The critical lattice modulation depth is a non-monotonic function of the lattice chirp (Fig. 1(d)). There exists an optimal chirp value at which $p_{cr}$ is minimal. For small $\alpha$ the transverse localization of linear mode increases as $\sim \alpha^{-1/2}$, while for high enough $\alpha$ delocalization due to the field coupling into channels in the lattice depth becomes significant and results in the formation of slowly decaying mode tails. From physical point of view, the key feature of such surface wave is that its profile forms as interference pattern of distributed Bragg-type reflection from the lattice interior counterbalanced by the total internal reflection of Fourier-components from the very interface.

The properties of the nonlinear waves supported by the interface with chirped lattices are summarized in Fig. 2. The energy flow

$$U = \int_{-\infty}^{\infty} |q|^2 \, d\eta \qquad (2)$$

is a monotonically increasing function of the propagation constant $b$ (Fig. 2(b)). When $U \to 0$ the surface soliton profile transforms into a linear guided mode whose maximum is located either in the near-surface lattice channel if $p > p_{\text{cr}}$, or shifts into the lattice for $p < p_{\text{cr}}$. Notice the absence of power threshold for formation of these surface waves at inter-faces, in contrast to interfaces with perfectly periodic lattices, where surface soliton formation is possible only for $U > U_{\text{th}}$. For comparison, in Fig. 2(b) we present $U(b)$ curves for both chirped and periodic lattice interfaces for similar values of the lattice modulation depth and carrying frequency. An increase of the energy flow, or of the propagation constant, is accompanied by progressive soliton localization in the near-surface guiding channel (Fig. 2(a)). Linear stability analysis indicates that surface solitons at chirped interfaces are stable in the entire domain of their existence.

One central result of this work is the attraction that the chirped interfaces cause on light beams. Thus, surface solitons can be excited even with the light beams launched in the lattice region remarkably far away from the very interface. An intuitive analysis of soliton dynamics can be carried out by means of the effective particle method. The effective potential $\Pi(\eta)$ for a soliton beam with the functional shape $\chi \operatorname{sech}[\chi(\eta - \eta_0)]$ and form-factor $\chi$ is defined by the instantaneous frequency of the chirped lattice $\Omega(\eta) = \Omega_0(1 + \alpha\eta)$. For $\eta \geq 0$ it takes the form $\Pi(\eta) = 2p[\pi\Omega(\eta)/2\chi]\{1 - \cos[\Omega(\eta)\eta]\}/\sinh[\pi\Omega(\eta)/2\chi]$, provided that the chirp parameter is sufficiently small ($\alpha \ll 1$). One can clearly see that the deepest well in effective potential is situated near the interface and the depths of subsequent potential wells diminish almost exponentially as one shifts into the lattice depth due to the factor $\sinh[\pi\Omega(\eta)/2\chi]$ present in the denominator. Therefore, on intuitive grounds one expects a tendency for solitons launched away from the interface to jump between neighboring wells until the localization in the deepest near-surface well is achieved. Therefore, the interface of chirped lattice may act as a *soliton attractor*. This is consistent with the physical properties of the refractive index landscape, as light tends to concentrate in regions with higher refractive index, i.e. in the vicinity of interface.

Figure 3 illustrates that such phenomenon does occurs. To analyze soliton dynamics we solved Eq. (1) with the input condition $q(\eta,0) = \chi \text{sech}[\chi(\eta - \eta_0)]$, where $\eta_0 = 5$. Notice that while the progressive transverse deflection of light beams in lattices with linear frequency or amplitude modulation was reported earlier [17], the presence of interface in our case dramatically modifies the dynamics of the beam propagation in the very vicinity of the interface and brings new specific features (such as a possibility of consecutive beam reflections) that are absent in infinite chirped lattices. Therefore, further we are interested in new effects that arise due to the interplay between interaction with interface and deflecting force exerted on solitons in chirped lattices, and their role in excitation of surface solitons. In low frequency lattices the input beam is deflected toward the surface due to the gradient of the effective potential $\Pi(\eta)$ and is quickly trapped in the near-the-surface channel, thereby exciting a surface soliton almost immediately after the first contact (Fig. 3(a)). Importantly, in high-frequency lattices both deflection rate and mobility of the beam increases, so that the beam may experience multiple reflections from the interface, but periodically returns to it due to the potential gradient. As a result, a process of multiple bouncing of a beam is accompanied by emission of radiation and gradually leads to the surface soliton formation (Fig. 3(b)). Figure 3(c) shows the dependence of the distance $\xi_c$ where beam collides with the interface for the first time on the lattice modulation depth $p$. The collision distance decreases monotonically with $p$, because the effective potential gradient grows with $p$. The complex influence of the chirp parameter $\alpha$ on the collision length is illustrated in Fig. 3(d). A well-defined minimum of collision distance at certain chirp value takes place. The presence of such minimum can be explained by the fact that at small values of $\alpha$ the deflection is not strong enough to cause soliton jumps between neighboring lattice channels, while at large values of $\alpha$ the broad beam smoothes over rapid oscillations of the effective potential and experiences almost no net attraction towards the interface.

Figure 4 displays the influence of the input beam power $U = 2\chi$ on the dynamics of surface wave formation. Increasing the input power initially enhances the deflection rate and decreases the characteristic distance of surface wave formation (Figs. 4(a) and 4(b)), but above a certain threshold energy level the input beam becomes immobile and remains in the lattice channel where it was launched (Fig. 4(c)). This is because the nonlinear contribution to the potential depth well becomes dominant in this regime.

Summarizing, we showed that semi-infinite optical lattices with a linear spatial frequency chirp, imprinted in focusing cubic media, support power thresholdless surface waves. The unique properties of such interfaces result in an attraction towards the interface for light beams launched inside the lattice, so that surface soliton formation becomes almost insensitive to the weak perturbations of the input beam profiles. The transverse localization length (or the width of the surface wave) might be effectively controlled by varying the modulation depth and frequency chirp of the lattice. Notice that thresholdless surface waves and attraction by the interface can occur also at the interfaces of lattices with amplitude modulation.

This work was partially supported by CONACyT under project 46522 and by the Government of Spain through grant TEC2005-07815/MIC and Ramon-y-Cajal program.

# References

1. Nonlinear surface electromagnetic phenomena, Ed. by H. E. Ponath and G. I. Stegeman, North Holland, Amsterdam (1991).
2. D. Mihalache, M. Bertolotti, and C. Sibilia, Progr. Opt. **27**, 229 (1989).
3. K. G. Makris, S. Suntsov, D. N. Christodoulides, and G. I. Stegeman, Opt. Lett. **30**, 2466 (2005).
4. S. Suntsov, K. G. Makris, D. N. Christodoulides, G. I. Stegeman, A. Haché, R. Morandotti, H. Yang, G. Salamo, and M. Sorel, Phys. Rev. Lett. **96**, 063901 (2006).
5. Y. V. Kartashov, V. A. Vysloukh, and L. Torner, Phys. Rev. Lett. **96**, 073901 (2006).
6. M. I. Molina, R. A. Vicencio, and Y. S. Kivshar, Opt. Lett. **31**, 1693 (2006).
7. C. R. Rosberg, D. N. Neshev, W. Krolikowski, A. Mitchell, R. A. Vicencio, M. I. Molina, and Y. S. Kivshar, Phys. Rev. Lett. **97**, 083901 (2006).
8. E. Smirnov, M. Stepic, C. E. Rüter, D. Kip, and V. Shandarov, Opt. Lett. **31**, 2338 (2006).
9. Y. V. Kartashov, V. A. Vysloukh, and L. Torner, Opt. Express **14**, 12365 (2006).
10. G. A. Siviloglou, K. G. Makris, R. Iwanow, R. Schiek, D. N. Christodoulides, G. I. Stegeman, Y. Min, and W. Sohler, Opt. Express **14**, 5508 (2006).
11. Y. V. Kartashov, V. A. Vysloukh, D. Mihalache, and L. Torner, Opt. Lett. **31**, 2329 (2006).
12. Y. V. Kartashov, L. Torner, and V. A. Vysloukh, Opt. Lett. **31**, 2595 (2006).
13. K. G. Makris, J. Hudock, D. N. Christodoulides, G. I. Stegeman, O. Manela, and M. Segev, Opt. Lett. **31**, 2774 (2006).
14. M. I. Molina, I. L. Garanovich, A. A. Sukhorukov, and Y. S. Kivshar, Opt. Lett. **31**, 2332 (2006).
15. K. Motzek, A. A. Sukhorukov, and Y. S. Kivshar, Opt. Lett. **31**, 3125 (2006).
16. A. A. Sukhorukov, D. N. Neshev, A. Dreischuh, R. Fischer, S. Ha, W. Krolikowski, J. Bolger, A. Mitchell, B. J. Eggleton, and Y. S. Kivshar, Opt. Express **14**, 11265 (2006).
17. Y. V. Kartashov, V. A. Vysloukh, and L. Torner, J. Opt. Soc. Am. B **22**, 1356 (2005).

18. Y. V. Kartashov, L. Torner, and V. A. Vysloukh, Opt. Express **13**, 4244 (2005).
19. T. Pertsch, U. Peschel, and F. Lederer, Chaos **13**, 744 (2003).
20. D. Kossel, J. Opt. Soc. Am. **56**, 1434 (1966).
21. P. Yeh, A. Yariv, and A. Y. Cho, Appl. Phys. Lett. **32**, 104 (1978).
22. Y. V. Kartashov and V. A. Vysloukh, Phys. Rev. E **72**, 026606 (2005).

## Figure captions

Figure 1 (color online). (a) Normalized dispersion curves for linear surface modes at $\alpha=0.05$. (b) Profiles of linear surface modes at $\Omega_0=2$, $\alpha=0.05$ corresponding to points marked by circles in (a). In white regions $R(\eta)<1$, while in gray regions $R(\eta)\geq 1$. (c) Critical lattice depth versus $\Omega_0$ at $\alpha=0.15$. (d) Critical lattice depth versus $\alpha$ at $\Omega_0=5$.

Figure 2. (a) Profiles of surface solitons at $\Omega_0=5$, $p=5$, $\alpha=0.1$. In white regions $R(\eta)<1$, while in gray regions $R(\eta)\geq 1$. (b) Energy flow versus propagation constant for solitons at interfaces with chirped and periodic lattices at $\Omega_0=5$, $p=5$. Points marked by circles correspond to profiles shown in (a).

Figure 3 (color online). Dynamics of excitation of surface waves at (a) $\Omega_0=3$, $p=1$, $\alpha=0.15$ and (b) $\Omega_0=6$, $p=4.5$, $\alpha=0.15$. (d) Collision distance versus lattice depth at $\Omega_0=3$, $\alpha=0.15$ (c) and versus lattice chirp at $\Omega_0=3$, $p=1$ (d). In all cases input soliton with $\chi=1$ was launched at $\eta=5$. Vertical dashed lines in (a) and (b) indicate interface position.

Figure 4 (color online). Propagation dynamics of solitons with form-factors $\chi=0.7$ (a), 0.9 (b), and 1.4 (c) in the vicinity of interface with chirped lattice. In all cases $\Omega_0=3$, $p=2$, $\alpha=0.15$ and soliton was launched at $\eta=5$.

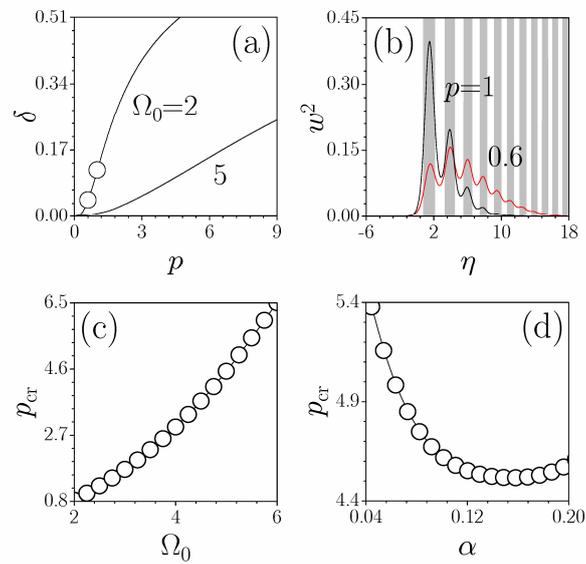

Figure 1 (color online). (a) Normalized dispersion curves for linear surface modes at $\alpha=0.05$. (b) Profiles of linear surface modes at $\Omega_0=2$, $\alpha=0.05$ corresponding to points marked by circles in (a). In white regions $R(\eta)<1$, while in gray regions $R(\eta)\geq 1$. (c) Critical lattice depth versus $\Omega_0$ at $\alpha=0.15$. (d) Critical lattice depth versus $\alpha$ at $\Omega_0=5$.

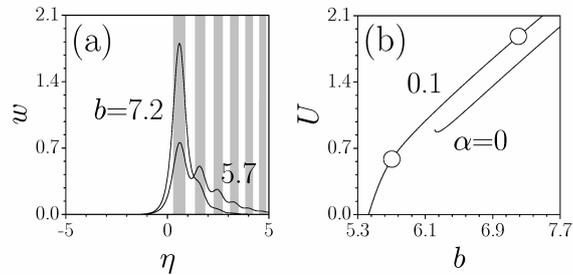

Figure 2. (a) Profiles of surface solitons at $\Omega_0 = 5$, $p = 5$, $\alpha = 0.1$. In white regions $R(\eta) < 1$, while in gray regions $R(\eta) \geq 1$. (b) Energy flow versus propagation constant for solitons at interfaces with chirped and periodic lattices at $\Omega_0 = 5$, $p = 5$. Points marked by circles correspond to profiles shown in (a).

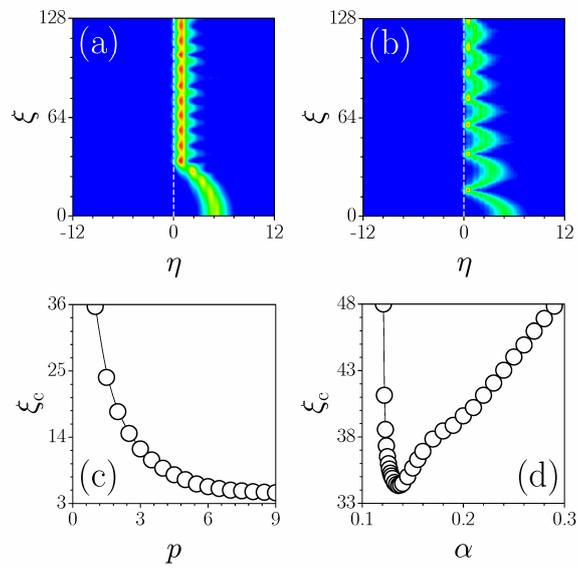

Figure 3 (color online). Dynamics of excitation of surface waves at (a) $\Omega_0 = 3$, $\alpha = 0.15$, $p = 1$ and (b) $\Omega_0 = 3$, $\alpha = 0.15$, $p = 5$. Collision distance versus lattice depth at $\Omega_0 = 3$, $\alpha = 0.15$ (c) and versus lattice chirp at $\Omega_0 = 3$, $p = 1$ (d). In all cases input soliton with $\chi = 1$ was launched at $\eta = 5$. Vertical dashed lines in (a) and (b) indicate interface position.

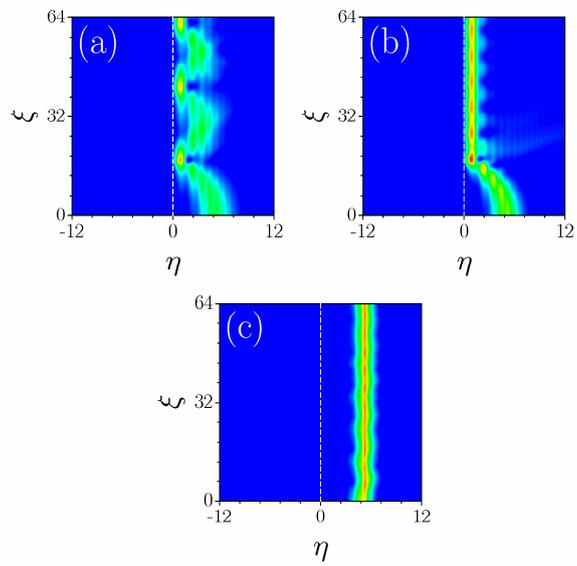

Figure 4 (color online). Propagation dynamics of solitons with form-factors $\chi = 0.7$ (a), $0.9$ (b), and $1.4$ (c) in the vicinity of interface with chirped lattice. In all cases $\Omega_0 = 3$, $p = 2$, $\alpha = 0.15$ and soliton was launched at $\eta = 5$.